**Abstract**

*This is a supplement to the paper arXiv:q-bio/0701050, containing the text of correspondence sent to **Nature** in 1990.*

**Origin of adaptive mutants: a quantum measurement?**

Sir, - Several recent works described non-random induction of adaptive mutations by environmental stimuli [1-3]. The most obvious explanation of this striking phenomenon would be that activation of gene expression leads to the enhancement of its mutation rate[4]. However, this does not work with the lacZ mutations described by Cairns and co-workers as the true inducer of the lac-operon is not lactose as such, but allolactose, a by-product of the β-galactosidase reaction[5]. So, in lacZ mutants the operon is not induced by lactose[6]. Besides, induction of respective genes would not explain the high fraction, among the revertants, of suppressor mutations in tRNA genes[1,7].

Other explanations suggest some special mechanisms for the "acceleration of adaptive evolution", like selection of "useful" protein coupled to specific reverse transcription[1]. However, any mechanism of this type also should have emerged in evolution. I propose that, to explain the adaptive mutation phenomenon, there is no need for any new *ad hoc* mechanism. The only thing that is necessary is to return to the old discussion of the role of quantum concepts in our understanding of life. This alone will allow the explanation of this manifestly Lamarckian phenomenon by Darwinian selection, occurring not in a population of organisms as usual, but in a "population" of virtual, in the direct quantum theory sense, states of each distinct cell. Thus, this hypothesis may be called "selection of virtual mutations". Detailed substantiation of this concept will be presented in a special publication; below I briefly show how this explanation might work.

It has been shown by the Cairns group that the mutations ensuring cell growth begin to accumulate not immediately after plating, but only after conditions are created under which such mutations become "useful", as if the mutations are induced by these conditions[1]. I suggest that, to explain this phenomenon, we should change our ideas about what a cell is, and consider not *actual* but *virtual* mutations. An important distinction of virtual mutations is that they do not accumulate with time in stationary cell, whereas the number of actual mutants would grow linearly from the moment of plating, and this would yield drastically different results in experiments like those shown in Fig. 3 of Ref. 1. Virtual mutations produce "delocalization" of the cell among different states, similarly to the delocalization of electron in physical space. However, for a virtual mutation to become an actual one, certain conditions are necessary, namely the possibility to grow, leading the system away irreversibly from the initial state. Such conditions arise when, for example, lactose is added to a plate with lacZ bacteria. Briefly, this is the essence of the proposed explanation.

What is a virtual mutation? The main cause of usual spontaneous mutations is the well-known base tautomerization[8] (having the *in vitro* frequency of about $10^{-4}$). Thus could we reduce 'virtual mutation' to such tautomerization? I believe that this view is not consistent with experiments, as it implies that the same rare tautomeric form should work both in transcription and in replication. If these processes are considered independent, we logically arrive to the leaky mutant, which was refuted by Cairns and coworkers[1]. Thus we need to postulate a correlation between the recognition of the tautomeric forms in transcription and in replication, making us to define "virtual mutation" as a certain state of the cell as a whole. Analogous reasoning is applicable to the "adaptive transpositions" discussed by Cairns. In other words, we consider the whole cell as a quantum system, with non-negligible nonlocality inherent in such systems. Most of all it resembles the systems of "generalized rigidity"[9], such as superfluid or superconducting states of matter, whose behavior is linked to quantum correlations; and I believe, similar correlations take place in the cell too.

I would like to emphasize that the proposed approach does not require detalization of molecular processes in the cell. Its main focus is the behavior of the cell as a whole. Similarly, to explain gyroscopic precession there is no need to consider interactions between elements inside the gyroscope; it's enough to know some motion invariants, defined by space-time symmetries.

Starting from this general view, one may express the above ideas using the operator formalism, and considering experiments conducted by Cairns as measurement of the cells' capability to propagate under given conditions. I suggest that the trait "ability to reproduce on lactose" (as an example) can be represented by an operator which one may designate "$\mathcal{Lac}$". Importantly, this new operator will act on the state $\Psi$ of the whole cell because the ability to reproduce is a property of the cell as a whole, and not of any part of it. Generally, "$\mathcal{Lac}$" will decompose this $\Psi$ into a superposition of some eigenfunctions. The components of this superposition are those functions that do not change upon the action of this operator, but are only multiplied by a constant. It reflects the essence of operator formalism in quantum theory, which chooses states *compatible* with given experimental conditions. There are three such eigenfunctions (I intentionally simplify the situation): $\psi_1$ corresponds to cell death, $\psi_2$ to the stationary state, and $\psi_3$ to the self-reproduction (that is the virtual mutation, in our case). Each function will enter the decomposition of $\Psi$ with a coefficient $c_i$ related to the probability of this or that outcome, i.e.:

$$\Psi = c_1 \psi_1 + c_2 \psi_2 + c_3 \psi_3, \quad \text{where } \Sigma |c_i|^2 = 1$$

By plating the cells on lactose agar we, in fact, begin to measure their ability to grow under these particular conditions. The rate of accumulation of lac revertants, i.e. the probability to obtain a cell in the mutant state, will correspond to $|c_3|^2$, being a small, but finite quantity, appearing, for example, due to base tautomerization. Here, the role of cell growth is dual: on the one hand, it is a factor of irreversibility amplifying the "quantum fluctuation", and on the other hand, it is a selection criterion, as each kind of virtual mutants capable of growth under these conditions can lead to colony formation. Another situation, i.e. glucose/valine agar, will be represented by another operator ($\mathcal{Val}$), which will decompose the same $\Psi$ function according to another basis, and Val$^r$ mutants will be obtained with certain rate. In fact, this is the essence of adaptive mutation phenomenon, where a particular condition induces emergence of respective mutants.

Thus, the proposed change of our view on the cell suggests that, in accord with quantum concepts, we are not dealing with the probability for a cell to mutate by itself, independent of experimental conditions. Rather, we are dealing with the probability *to observe* the cell in the mutant state by plating it on lactose. We are certainly simplifying situation, as spontaneous mutations that accumulate during cell growth before plating, make our ensemble 'mixed'. However, this complication does not change the essence of the explanation, according to which adaptive mutations emerge by measurement of 'pure' state. This resembles the passage of a polarized photon through a polarizer turned under some angle to the photon polarization. It will be incorrect to say that the polarization of the photon could turn by the necessary angle by chance, prior to interaction. It is the specific experimental situation that makes us to decompose the state vector according to the respective basis states, and to evaluate the fraction of the component that will pass through polarizer. On the other hand, one may speak about "adaptation" of photon polarization by selection of "fit" eigenstate, and consider this case as the model for our phenomenon.

How are all these ideas applicable to the living bacterial cell? Discussion of the possible role of quantum concepts in biology has a rather long history, initiated by Niels Bohr ('the complementarity principle'). Briefly, one might reduce the essence of this discussion to the principal impossibility to predict precisely the fate of an individual cell. For example, any attempt to determine, whether it is able to reproduce under certain conditions, will lead to irreversible change of the state of the cell, even to its death. This is reminiscent of the two-slit diffraction experiment, where an attempt to determine through which of the two slits the electron actually passes will lead to disappearance of the interference. The two trajectories of the electron can be made physically discernable only by the cost of changing the experimental situation. Similarly, the notorious phenomenon of the "wholeness" of the living organism can be formally expressed according to the Feynman rules of calculating probabilities: different indiscernible (in the given experimental conditions) variants should be included in the pure state (i.e. their amplitudes, and not probabilities, should be summed, leading to interference and other quantum effects). Thus, as long as a whole cell exists and is alive, we are obligated to treat its different indiscernible states in this way. Such consideration of operational limitations allows us to explain the adaptive mutation phenomenon (and hopefully other adaptations too) as the consequence of unavoidable quantum scatter in measurement of the cell's capability to propagate under given conditions.

In spite of its apparent formal character, this hypothesis allows us to make some predictions of applied (in particular, medical) interest. It predicts that in processes involving somatic mutations (e.g. oncogenesis, or specific antibody generation) the mutations may be induced by conditions allowing the cell that happened to be in the state of virtual mutation to proliferate irreversibly. I believe, this possibility can be tested experimentally.

*Comments:*

This text was written in 1990. The author translated it to English with the kind help of Dr. Eugene Koonin (current affiliation: National Center for Biotechnology Information, National Library of Medicine, National Institutes of Health, Bethesda MD, USA.)

The English version of the text was sent to *Nature* in 1990 and rejected. At the same time it was also sent to the following correspondents :

1. JOHN CAIRNS
Department of Cancer Biology, Harvard School of Public Health, Boston, Massachusetts 02115.

2. BARRY HALL
Department of Molecular and Cell Biology, University of Connecticut, Storrs 06269.

3. BERNARD DAVIS
Bacterial Physiology Unit, Harvard Medical School, Boston, MA 02115.

4. KOICHIRO MATSUNO
Department of BioEngineering, Nagaoka University of Technology, Japan.

5. KONSTANTIN CHUMAKOV
Center for Biologics Evaluation and Research, Food and Drug Administration, Rockville, Maryland 20852, USA.

6. MIKHAIL V. IVANOV
Institute of Microbiology, Russian Academy of Sciences, pr. 60-letiya Oktyabrya 7, k. 2, Moscow, 117811 Russia.

, as well as to all participants of the discussion 'Origin of mutants disputed' (*Nature* **336**, 525 - 526 (08 December 1988)) :

1. D. CHARLESWORTH, B. CHARLESWORTH & J. J. BULL
Department of Ecology and Evolution, University of Chicago, 915 East 57th Street, Chicago, Illinois 60637, USA
Department of Zoology, University of Texas, Austin, Texas 78712, USA

2. ALAN GRAFEN
Animal Behaviour Research Group, Zoology Department, Oxford University, Oxford OX1 3PS, UK

3. R. HOLLIDAY & R. F. ROSENBERGER

CSIRO Laboratory for Molecular Biology, North Ryde, Sydney, Australia
Genetics Division, National Institute for Medical Research, Mill Hill, London NW7 1AA, UK

4. LEIGH M. VAN VALEN

Department of Ecology and Evolution, University of Chicago, 915 East 57Street, Chicago,
Illinois 60637, USA

5. ANTOINE DANCHIN

Institut Pasteur, 28 Rue Dr. Roux, 75724 Paris, Cedex 15, France

6. IRWIN TESSMAN

Departments of Biiological Sciences, Purdue University, West Lafayette,
Indiana 47907, USA